\documentstyle[12pt]{article}
\newcommand\be{\begin{equation}}
\newcommand\ee{\end{equation}}
\newcommand\bea{\begin{eqnarray}}
\newcommand\eea{\end{eqnarray}}
\newcommand\ket[1]{|#1\rangle}

\newcommand\braket[2]{\langle #1|#2\rangle}
\newcommand{\fatalpha}{{\bf \alpha \kern -0.44em \alpha}}
\newcommand{\fatsigma}{{\bf \sigma \kern -0.54em \sigma}}
\newcommand{\tpchi}{{\bf \chi \kern -0.35em \chi}}
\newcommand{\llambda}{{\bf \lambda \kern -0.45em \lambda}}

\bibliography{plain}
\title{\bf Continuous-Time Classical and Quantum Random Walk on Direct Product of Cayley Graphs  }\vspace{20mm}
\author{ S. Salimi$^{a}$
  \thanks{Corresponding author:  E-mail addresses: shsalimi@uok.ac.ir},
  M. A. Jafarizadeh$^{b,c,d}$
 \thanks{E-mail: jafarizadeh@tabrizu.ac.ir}
 \\ $^a${\small Department of Physics,
University of Kurdistan, Kurdistan 51664, Iran.}\\  $^b${\small
Department of Theoretical Physics and Astrophysics, Tabriz
University, Tabriz 51664, Iran.} \\ $^c${\small Institute for
Studies
in Theoretical Physics and Mathematics, Tehran 19395-1795, Iran.} \\
$^d${\small Research Institute for Fundamental Sciences, Tabriz
51664, Iran.}}  \pagebreak

\vspace{20mm}
\begin{document}
\maketitle \vspace{15mm}
\newpage
\begin{abstract}
In this paper we define direct product of graphs and give a recipe
 for obtained probability of observing particle on vertices in the
 continuous-time classical and quantum random walk.
In the recipe,  the probability of observing particle on direct
product of graph obtain  by multiplication of probability on the
corresponding to  sub-graphs, where this method is useful  to
determine probability of walk on complicated graphs. Using this
method, we calculate the probability of continuous-time classical
and quantum random walks on many of finite direct product cayley
graphs ( complete cycle, complete $K_n$, charter and $n$-cube).
Also, we inquire that the classical state the stationary uniform
distribution is reached as $t\longrightarrow \infty$ but for quantum
state is not always satisfy.


{\bf PACs Index: 03.65.Ud }
\end{abstract}
\vspace{70mm}
\newpage
\section{Introduction}

The theory of Markov chains and random walks on graphs is
fundamental to mathematics, physics, and computer science
\cite{d}, \cite{ll}, \cite{rmr} as it provides a beautiful
mathematical framework to study stochastic process and its
applications. Among some of the known examples of these
applications include Monte Carlo methods in statistics, the theory
of diffusion in statistical physics, and algorithmic techniques
for sampling and random generation of combinatorial structures in
computer science(based on rapid mixing of certain Markov chains).
 Two pervasive algorithmic ideas in quantum computation are
Quantum Fourier Transform (QFT) and amplitude amplification (see
\cite{cn}). Most subsequent progress in quantum computing owed
much to these two beautiful ideas. But there are many problems
whose characteristics matches neither the QFT nor the amplitude
amplification mold (e.g., the Graph Isomorphism problem). This
begs for new additional tools to be discovered.

  A natural way to discover new quantum algorithmic ideas is to adapt
a classical one to the quantum model. An appealing well-studied
classical idea in statistics and computer science is the method of
random walks \cite{diaconis, motwani}. Recently, the quantum
analogue of classical random walks has been studied in a flurry of
works \cite{fg,cfg,abnvw,aakv,mr,k}. The works of Moore and Russell
\cite{mr} and Kempe \cite{k} showed faster bounds on instantaneous
mixing and hitting times for discrete and continuous quantum walks
on the hypercube (compared to the classical walk). A recent work by
Childs et al. \cite{ccdfgs} gave an interesting and powerful
algorithmic application of continuous-time quantum walks.

A study of quantum walks on simple lattice is well known in
physics(see \cite{fls}). Recent studies of quantum walks on more
general graphs were described in \cite{cfg,fg,aakv,ccdfgs,js,jsa,
jsas,sa1, sa2, konno1, konno2,adz}. Some of these works studies the
problem in the important context of algorithmic problems on graphs
and had suggested that quantum walks is a promising algorithmic
technique for designing future quantum algorithms.

Several important classes of graphs studied in classical random
walks include the binary $n$-cube, the circulant graphs, and the
group-theoretic Cayley graphs. The binary $n$-cube and circulant
graphs are important in the study of interconnection networks and
complexity of Boolean function, and Cayley graphs capture strong
gorup-theortic ingredients of important problems, such as Graph
Isomorphism. sine most of these graphs are regular, classical random
walks on them are known to converge or to mix towards the uniform
stationary distribution. The mixing properties of continuous-time
quantum walks on the same graphs were found to exhibit non-classical
behavior \cite{mr,abtw,aaht,gw}.

 In this paper we define direct product of graphs and give a recipe
 for obtained probability of observing particle on vertices in the
 continuous-time classical and quantum random walk.
In the recipe  the probability of observing particle on direct
product of graph obtain  by multiplication of probability on the
corresponding to  sub-graphs. This method is useful  to determine
probability of walk on complicated graphs. In the classical state
the stationary uniform distribution is reached as $t\longrightarrow
\infty$ but for quantum state is not satisfy. Using this method, we
calculate the probability of continuous-time classical and quantum
walk on many of finite direct product cayley graphs ( complete
cycle, complete $K_n$, charter and $n$-cube).

 The organization of this paper is as follow. In section
2, we give a brief outline of  graphs and their adjacency matrix.
 In section 3, we present a
concept of direct product of Cayley graphs. In the section $4$,
continuous-time classical random walks on graphs is studied, and we
calculate some examples. In the section $5$, we present
continuous-time quantum walk on graphs, and calculate some examples.
Finally, in section 6, the conclusions and future research are
presented.

\section{Graphs and its adjacency matrix}
Any mathematical object involving point and connections between them
may be called a graph. If all the connection are unidirectional, it
is call a digraph. A graph $\Gamma=(V,E)$ consists of two sets $V$
and $E$. The element of $V$ are called vertices (or node) and the
element of $E$ are called edges where is a subset of $\{\{x,y\}|x,
y\in V,x\neq y \}$. Two vertices $x, y\in V$ are called adjacent if
$\{\{x,y\}\in E$, and in this case we write $x\sim y$. We let $A$ be
the $|V|\times |V|$ adjacency matrix of $\Gamma$, i.e, $A$ is
indexed by elements of $V$ an is as follow:
\begin{equation}
A_{xy} = \left\{
\begin{array}{ll}
1 & \mbox{if $x\sim y$}\\
0 & \mbox{otherwise}
\end{array}
\right.
\end{equation}

Obviously, (i) $A$ is symmetric; (ii) an element of $A$ takes a
value in $\{0, 1\}$; (iii) a diagonal element of $A$ vanishes.
Conversely, for a non-empty set $V$, a graph structure is uniquely
determined by such a matrix indexed by $V$. The \emph{degree} or
\emph{valency} of a vertex $x\in V$ is defined by
\begin{equation}
\kappa(x)=|\{y\in V; x\sim y\}|,
\end{equation}
where $\mid.\mid$ denotes the cardinality. A finite sequence $x_0;
x_1; . . . ; x_n \in V$ is called a walk of length $n$ (or of $n$
steps) if $x_{k-1}\sim x_k$ for all $k=1, 2, . . . , n$. In a walk
some vertices may occur repeatedly.

\section{The Direct product of cayley  Graphs }
In this section, we briefly discuss necessary background information
on Cayley graphs of the group and will study their product.

 Let $\Gamma_i, \; i=1,2,..,d$ be
graphs of finite vertices with the corresponding adjacency matrices
$A_i, i=1,2,...,d$. Then their direct product
 \be\label{p1}\label{gp}
\Gamma_1\otimes\cdots \otimes \Gamma_d,
 \ee
is a graph with the following  adjacency matrix  $A$:
  \be\label{a1}
  A=\sum_{j=1}^{d}
  I\otimes\cdots\otimes A_j\otimes\cdots\otimes I
  \ee
where the $j$th term in the sum has $A_j$ appearing in the $j-th$
place in the tensor product.

Let $G$ be a finite group and let $R\subseteq G$ be a set of
generators for $G$ satisfying $g\in R\Leftrightarrow g^{-1}\in R$
for all $g\in G$ (in this case the Cayley graph will be equivalent
to an undirected regular graph). Then the Cayley graph of $G$ with
respect to $R$, which we denote by $\Gamma(G,R)$ in this paper, is
an undirected graph defined as follows. The set of vertices of
$\Gamma(G,R)$ coincides with $G$, and for any $g,h\in G$, $\{g,h\}$
is an edge in $\Gamma(G,R)$ if and only if $g h^{-1}\in R$.

Now let $\Gamma_{i}$ be Cayley graph of finite group $G_{i}$, with
respect to $R_i$. Then the graph  $\Gamma$ is generated by using
Eq.(\ref{gp}) is the Cayley graph of finite group
$G=G_1\otimes\cdots\otimes G_n$,
  with respect to
  $R=\{(r_{j}^{(1)},1,...,1),(1,r_{j}^{(2)},1,...,1),...,(1,1,...,r_{j}^{(n)})\}$,
   where $r_{j}^{(i)}\in R_i$ and $1$ as the neutral element of $G_i$  .
Thus, for any $g=(a_1,...,a_n), h=(b_1,...,b_n) $ where $a_i, b_i
\in G$, the connected is defined if $gh^{-1}\in R$ ,i.e,
$(a_1b^{-1}_1,...,a_{n}b^{-1}_n)\in R$, consequently, $a_i\neq b_i$
only in one element (the vertices $a_{i_{1},i_{2},i_{3},\cdots
,i_{n}}$ and
 $a_{j_{1},j_{2},j_{3},\cdots,
 j_{n}}$ are connected provided that they differ only in one
 indices, i.e, $i_{k}=j_{l}$ for $k,l=1,2,...,m-1,m+1,...,n$ but
 $i_{m}\neq j_{m}$).

In the end we argue circulant graph which to deem necessary in the
examples that we will investigate continuous-time classical and
quantum random walk for them.

If $G$ is a cyclic group, then the Cayley graph is called a
circulant graph. The adjacency matrix $A$ of circulant graph is
given by
\begin{equation}
A=\sum_{k=0}^{n-1}a_{k} P^k,
 \end{equation}
where $P$ is the $n\times n$ primary permutation matrix\cite{agr}
as follow :
\begin{equation}
P=\left(\begin{array}{cccccc}0 & 1 & 0 & 0 & \cdots & 0
\\0 & 0 & 1 & 0 & \cdots & 0\\
\vdots & \vdots & \vdots & \vdots & \vdots & \vdots
\\ 0 & 0 & 0 & 0 & \cdots & 1\\
1 & 0 & 0 & 0 & \cdots & 0
\end{array}\right).
\end{equation}
The adjacency matrix $A$ is belonging to $CG-$modules, that $C$ and
$G=<P>$ are complex field and cyclic group with $ord(P)=n$,
respectively. A well-known theorem states that any $CG$-modules can
be expressed as a direct sum of irreducible $CG$-submodules. Also,
the fact that the dimensional of every irreducible $CG$-modules is
one for finite abelian group $G$, and the dual group of a cyclic
group is isomorphic to the group itself, can be used Fourier
transformation for diagonalizable of adjacency matrix $A$.

Thus the primary permutation matrix is that diagonalizable
(unitarily) by the Fourier matrix
\begin{equation}\label{Cir}
    F = \frac{1}{\sqrt{n}}V(\omega),
\end{equation}
where $\omega = e^{2\pi i/n}$ and $V(\omega)$ is the Vandermonde
matrix. Therefore, we have
\begin{equation} \label{eqn:circ}
    F^{\dagger}PF = diag(1,\omega,\omega ^2,\omega ^3,\omega ^4,...,\omega ^{n-1}).
\end{equation}
This Equation (\ref{eqn:circ}) shows that the eigenvalues of a
circulant matrix can be obtained by using the Fourier transform
$F$.

 Some examples of circulant graphs that we study, they are the
complete graphs(i.e, $a_1=a_2=\cdots a_{n-1}=1$, $a_0=0$, $G=Z_N$
and $R=\{1,2,...,N-2\}$) and full-cycle(i.e, $a_1=a_{n-1}=1$ and
otherwise $0$, $G=Z_N$ and $R=\{1,N-2\}$ ).

Also, two other examples  , for direct product graphs that we study,
are the hypercubes simple structure as a product simplex graph
$K_2$, and charter graph(i.e, $G=Z_2\otimes Z_n $ such that the
corresponding product graph's is $\Gamma=K_2\otimes C_n$ ).

\section{Continuous-time classical random walks on direct product of graphs}
Let $\Gamma=(V,E)$ be a simple (no self-loops), undirected,
connected graph with adjacency matrix $A$. Suppose that
$P:V(\Gamma)\longrightarrow [0,1]$ denotes a time-dependent
probability distribution of a stochastic process on $\Gamma$. The
classical evolution of the discrete-time random walk is given by the
equation
\begin{equation}
P(t)=W^{t}P(0),
\end{equation}
where $W$ is the stochastic transition matrix and $t\in Z^{+}$. In a
simple walk on a $d$-regular graph $\Gamma$, we let
$W_{d}=\frac{1}{d}A$; this defines a random walk where, at each
step, the particle moves to one of its $d$ neighbors randomly. On
the other hand, in a \emph{lazy} walk on $\Gamma$, the particle
stays or moves to a random neighbor with equal probabilities; here
we have $W_{l}=\frac{1}{2}+\frac{1}{2}W_{d}$.

The Laplacian of $\Gamma$ is defined as $H=A-D$, where $D$ is a
diagonal matrix whose $j$-th entry is the degree of vertex $j$ of
$G$. Suppose that $P(t)$ is a probability distribution of
continuous-time walk at time $t$. The classical evolution of the
continuous-time walk is given by the Kolmogorov equation
\begin{equation}
\frac{d}{dt}P(t)=HP(0).
\end{equation}
The solution of this equation, modulo some conditions, is
\begin{equation}
P(t)=e^{tH}P(0).
\end{equation}
Thus, the solution for the product of cayley graphs
equation(\ref{p1}) with the normalized adjacency matrix
equation(\ref{a1}) such that $H_i=A_{d_i}-D_{d_i}$ is as follow :
\begin{equation}
P(t)=\prod_{i=1}^{d}\otimes(e^{(A_{n_i}-D_{n_i})t} )P(0),
\end{equation}
with initial probability $P(0)= P_1(0)\otimes\cdots\otimes P_d(0)$.
Then we obtain probability distribution of continuous-time walk at
time $t$ as
\begin{equation}
P(t)=\prod_{i=1}^{d}\otimes(e^{(A_{n_i}-D_{n_i})t}
P_i(0))=\prod_{i=1}^{d}\otimes (P_i(t)).
\end{equation}

Therefore the probability for observing particle on direct product
of graph obtain  by multiplication of probability on the sub-graphs.
This method is useful  to determine probability of walk on
complicated graphs.

\subsection{The cycle graph $C_n$}
 Let $\frac{1}{2}A_n$ be the normalized adjacency
matrix of the full-cycle $C_n$ on $n$ vertices. According to
\cite{abt}, let $H=\frac{1}{2}A_{n}-I_{n}$ be its Laplacian matrix
of the full-cycle. Thus, the eigenvalues of $H$ is given by,
$\lambda _{j}=\cos(\frac{2\pi j}{n})-1$.

 The direct product for cycle graphs
is $C_{n_1}\otimes\cdots\otimes C_{n_d}$, such that the
corresponding Laplacian as follow :
 \begin{equation}
A=\sum_{i=1}^{d} I_{n_1}\otimes\cdots\otimes
(\frac{1}{2}A_{n_i}-I_{n_i})\otimes\cdots\otimes I_{n_d}
\end{equation}
where the $i$th term in the sum has $(\frac{1}{2}A_{n_i}-I_{n_i})$
appearing in the $i$th place in the tensor product. In order to, The
solution of  Kolmogorov equation for the direct product finite cycle
is
\begin{equation}
P(t)=\prod_{i=1}^{d}\otimes(e^{(\frac{1}{2}A_{n_i}-I_{n_i})}
)P(0).
\end{equation}

Using the orthonormal eigenvectors
$\ket{F_{j}}=\frac{1}{\sqrt{n}}\ket{\omega_{j}}$ (the columns of the
Fourier matrix F (\ref{Cir})) and the initial probability vector
$P(0)= P_1(0)\otimes\cdots\otimes P_d(0)$, then for $P(t)$, We have
\begin{equation}
P(t)=\prod_{i=1}^{d}\otimes(\frac{1}{n_i}\sum_{j=0}^{n_i-1}e^{t(-1+
    \cos(2\pi j/n_i))}
    \ket{\omega_j}_i)
\end{equation}

Thus for calculate the probability of particle, we use the
probability of particle on single graph $C_{n}$(i.e,
$P_{s,k}(t)=\frac{1}{n}\sum_{j=0}^{n-1}e^{t(1-\cos(2\pi j/n))}
    \omega_{j}^k$).
Therefore, the probability for observing the particle at the
position $\vec{K}$ is
\begin{equation}
P_{\vec{K}}(t)=\prod_{i=1}^{d} P_{s,k_i}(t).
\end{equation}
So, the stationary uniform distribution is reached as
$t\longrightarrow \infty$. The especially, if $n_1=n_2=\cdots
=n_d=n$, we have
\begin{equation}
P(t)=(e^{(\frac{1}{2}A_{n}-I_{n})})^{\otimes
d}P(0)=(\frac{1}{n}\sum_{j=0}^{n-1}e^{t(-1+
    \cos(2\pi j/n))}\ket{\omega_j})^{\otimes d}.
\end{equation}
.

\subsection{The complete graph $K_n$}
Let $\frac{1}{n-1}A_n$ be the normalized adjacency matrix of the
complete graph $K_n$ on $n$ vertices. Thus, $H=\frac{1}{n-1}A_n-I_n$
is the Laplacian matrix of the complete graph $K_n$. The eigenvalues
of $H$ are $0$ (once) and $-\frac{n}{n-1}$ (n-1 times). The direct
product of complete graphs is $K_{n_1}\otimes\cdots\otimes K_{n_d}$,
such that the corresponding Laplacian as follow :
\begin{equation}
A=\sum_{i=1}^{d} I_{n_1}\otimes\cdots\otimes
(\frac{1}{n_i-1}A_{n_i}-I_{n_i})\otimes\cdots\otimes I_{n_d}
\end{equation}
where the $i$th term in the sum has
$(\frac{1}{n_i-1}A_{n_i}-I_{n_i})$ appearing in the $i$th place in
the tensor product. The solution of Kolmogorov equation for direct
product complete graphs is
\begin{equation}
P(t)=\prod_{i=1}^{d}\otimes(e^{(\frac{1}{n_i-1}A_{n_i}-I_{n_i})}
)P(0)
=\left(\begin{array}{c} \frac{1}{n_1}(1+e^{-\frac{n_{1}t}{n_{1}-1}})\\
\frac{1}{n_1}(1-e^{-\frac{n_{1}t}{n_{1}-1}})\\ \vdots \\
\frac{1}{n_1}(1-e^{-\frac{n_{1}t}{n_{1}-1}})\\
\end{array}\right)\otimes\cdots\otimes\left(\begin{array}{c} \frac{1}{n_d}(1+e^{-\frac{n_{d}t}{n_{d}-1}})\\
\frac{1}{n_d}(1-e^{-\frac{n_{d}t}{n_{d}-1}})\\ \vdots \\
\frac{1}{n_d}(1-e^{-\frac{n_{d} t}{n_{d}-1}})\\
\end{array}\right)
\end{equation}
with initial probability $P(0)= P_1(0)\otimes\cdots\otimes P_d(0)$.
Thus the probability of particle at the position
$\ket{\vec{K}}=\ket{i_1,i_2...i_d}$,  is as follow :
\begin{equation}
 P_{\vec{K}}(t)=\prod_{l=1}^{d}
\frac{1}{n_l}[(1+e^{-\frac{n_{l}t}{n_{l}-1}})\delta_{i_l,0}+(1-e^{-\frac{n_{l}t}{n_{l}-1}})(1-\delta_{i_l,0})].
\end{equation}
So, the stationary uniform distribution is reached as
$t\longrightarrow \infty$.

The especially, if $n_1=n_2=\cdots=n_d=n$, we have
\begin{equation}
P(t)=(e^{(\frac{1}{n-1}A_n-I_{n})})^{\otimes d}P(0)
=\left(\begin{array}{c} \frac{1}{n}(1+e^{-\frac{nt}{n_{1}-1}})\\
\frac{1}{n}(1-e^{-\frac{nt}{n-1}})\\ \vdots \\
\frac{1}{n}(1-e^{-\frac{nt}{n-1}})\\
\end{array}\right)^{\otimes d}
\end{equation}
with initial probability $P(0)= P_1(0)\otimes\cdots\otimes P_d(0)$.
Thus for calculate the probability of particle, we use the
probability of particle on single graph $K_{n}$(i.e, $P_{s,0}(t)=
\frac{1}{n}(1+e^{-\frac{nt}{n-1}}),
P_{s,j}(t)=\frac{1}{n}(1-e^{-\frac{nt}{n-1}})$ for $j\neq 0$).
Therefore, the probability for observing the particle at the
position $\vec{K}$ is
\begin{equation}
P_{\vec{K}}(t)=(P_{s,0})^k (P_{s,j})^{d-k}
\end{equation}
where the $k$ is the number of zeroes.

\subsection{Charter}
As an example axiomatic, direct product graphs for the
continuous-time random walk, we can calculate the probability $P(t)$
directly by exploiting the charter's simple structure as a product
simplex graph $K_2$ and cycle graph $C_n$. Let $G=Z_2\otimes Z_n$ be
a complete 2-partite graph where each partition has $n>2$
vertices(the case $n=2$ is the square graph). The direct product of
complete graphs $K_2$ and cycle graph $C_n$ is $K_2\otimes C_n$ such
that the corresponding Laplacian as follow :
\begin{equation}
A=I_2\otimes(\frac{1}{2}A_n-I_n)+(\sigma_x-I_2)\otimes I_n .
\end{equation}
Thus, the solution of Kolmogorov equation for charter is
 $$ P(t)=e^{tA}P(0)=(e^{t(\sigma_x-I_2)}\otimes
e^{t(\frac{1}{2}A_n-I_n)})P(o)$$
\begin{equation}
=\frac{1}{2n}\left(\begin{array}{c}
(1+e^{-2t})\sum_{j=0}^{n-1}e^{t(-1+
    \cos(2\pi j/n))}\ket{\omega_j}\\
    (1-e^{-2t})\sum_{j=0}^{n-1}e^{t(-1+
    \cos(2\pi j/n))}\ket{\omega_j}
    \end{array}\right)
\end{equation}
Where, for calculate the probability of particle we use the
probability of particle on single graphs $K_2$ and $C_n$,
 with the initial probability $P(0)=P_1(0)\otimes P_2(0)$. So, the
 stationary uniform distribution is reached as $t\longrightarrow
 \infty$.

\subsection{$n$-cube}
As another example, direct product graphs for the continuous-time
random walk, we can calculate the probability $P(t)$ directly by
exploiting the hypercube's simple structure as a product simplex
graph. The binary $n$-cube is define over the set ${(0, 1)}^n$ of
$n$-cube binary sequence, where two sequence $x$ and $y$ are
connected if they differ exactly in one bit position. As it turns
out, the $n$-cube is also a complet graph.

 Let $\sigma_{x}$ (Pauli matrix) be the
normalized adjacency matrix of $K_2$, thus, $H=\sigma_{x}-I_2$ be
its Laplacian.  The direct product for hypercubes is
$S_{n}=K_{2}\otimes\cdots\otimes K_{2}$, such that the corresponding
Laplacian as follow :
\begin{equation}
A=\sum_{i=1}^{n} I_{2}\otimes\cdots\otimes
(\sigma_{x}-I_2)\otimes\cdots\otimes I_{2}
\end{equation}
where the $i$th term in the sum has $(\sigma_{x}-I_2)$ appearing
in the $i$th place in the tensor product. The solution of
Kolmogorov equation for hypercubes is
\begin{equation}
P(t)=(e^{(\sigma_{x}-I_{2})} )^{\otimes
d}P(0)=\left(\begin{array}{c} \frac{1}{2}(1+e^{-2t})\\
\frac{1}{2}(1-e^{-2t})\end{array}\right)^{\otimes d}
\end{equation}
with initial probability $P(0)= P_1(0)\otimes\cdots\otimes P_d(0)$.
Thus for calculate the probability of particle, we use the
probability of particle on single graph $Z_{2}$ (i.e, $P_{s,0}(t)=
\frac{1}{2}(1+e^{-2t}), P_{s,1}(t)=\frac{1}{2}(1-e^{-2t})$).
Therefore, the probability for observing the particle at the
position $\vec{K}$ is
\begin{equation}
P_{\vec{K}}(t)=\prod_{i=1}^{2} P_{s,k_i}(t).
\end{equation}
So, the stationary uniform distribution is reached as
$t\longrightarrow \infty$.

\section{Continuous-time quantum walks on direct product of graphs}
A continuous-time quantum walk is defined by replacing
Kolmogorov's equation with Schr\"{o}dinger's equation.
 Continuous-time quantum walks was introduced by Farhi and
Gutmann \cite{fg} (see also \cite{cfg,mr}). Our treatment, though,
follow closely the analysis of Moore and Russell \cite{mr} which we
review next. Let $\ket{\psi} : V(\Gamma)\longrightarrow C$ be a
time-dependent amplitude of the quantum process on $\Gamma$. The
wave evolution of the quantum walk is
\begin{equation}
    i\hbar\frac{d}{dt}\ket{\psi_{t}} = H\ket{\psi_{t}},
\end{equation}
where assume $\hbar = 1$,  and $\ket{\psi_{0}}$ be the initial
amplitude wave function of the particle, the solution is given by
$\ket{\psi_{t}} = e^{-iHt} \ket{\psi_{0}}$.

  But on $d$-regular graphs, $D = \frac{1}{d}I$, and since $A$ and $D$
commute, we get
\begin{equation} \label{eqn:phase-factor}
e^{-itH} = e^{-it(A-\frac{1}{d}I)} = e^{-it/d}e^{-itA}
\end{equation}
which introduces an irrelevant phase factor in the wave evolution.

The probability that the particle is at vertex $j$ at time $t$ is
given by
\begin{equation} \label{eqn:collapse}
    P_t(j) = |\braket{j}{\psi_{t}}|^2.
\end{equation}
The \emph{average} probability that the particle is at vertex $j$
is given by
 \be P(j)=\lim_{T\rightarrow\infty}
\frac{1}{T}\int_{0}^{T} P_{t}(j)dt.
 \ee

Since $H$ is Hermitian, the matrix $U_t = e^{-iHt}$ is unitary.
If $(\lambda_j,\ket{z_j})_j$ are the eigenvalue and eigenvector
pairs of $H$, then $(e^{-i\lambda_{j}t},\ket{z_j})_j$ are the
eigenvalue and eigenvector pairs of $U_t$. Because $H$ is
symmetric, there is an orthonormal set of eigenvectors, say
$\{\ket{z_j} : j \in [n]\}$ (i.e., $H$ is unitarily
diagonalizable). So, if $\ket{\psi_0} = \sum_j \alpha_j
\ket{z_j}$ then
\begin{equation}
    \ket{\psi_t} = \sum_j \alpha_{j}e^{-i\lambda_{j}t}\ket{z_j}.
\end{equation}
Hence, in order to analyze the behavior of the quantum walk, we
follow its wave-like patterns using the eigenvalues and
eigenvectors of the unitary evolution $U_t$. To observe its
classical
behavior, we collapse the wave vector into a probability vector using Equation (\ref{eqn:collapse}).\\
Thus, the wave function $\ket{\psi_t}$ for the product Cayley
graphs Eq.(\ref{p1}) with the normalized adjacency matrix
Eq.(\ref{a1}) such that $H_j=A_{n_j}-\frac{1}{n_j}I_{n_j}$
(regular graph) as follow :
\begin{equation}
\ket{\psi_t}=exp(-iAt)\ket{\psi_0}=e^{-itA_{n_1}}\ket{0}_1\otimes\cdots\otimes
e^{-itA_{n_d}}\ket{0}_d=\prod_{j=1}^{d} \otimes
e^{-itA_{n_j}}\ket{0}_{j}
\end{equation}
with initial state $\ket{\psi_0}=\ket{0}_1\cdots \ket{0}_d$. Also,
the above equation show that the  amplitude wave function of the
particle on direct product of graph obtain by multiplication of
amplitudes  at the sub-graphs. This method is useful  to determine
amplitudes of walk on complicated graphs.

\textbf{Definition 1} (\emph{instantaneous and average mixing } \cite{cfg,mr})\\
 \emph{ Let $\epsilon\geq 0$. A graph $G=(V,E)$ has the
 instantaneous $\epsilon$-uniform mixing property if there exists $t\in R^{+},$ such that
  the continuous-time quantum walk on $G$ satisfies $\parallel
  P_{t}-U\parallel\leq \epsilon$, where $\parallel
  Q_{1}-Q_{2}\parallel =\sum_{x}|Q_{1}(x)-Q_{2}(x)|$ is the total
  variation distance between two probability distribution
  $Q_{1},Q_{2},$ and $U$ us the uniform distribution on the
  vertices of $G$. Whenever $\epsilon =0$ is achievable, $G$ is said to have instantaneous exactly
  uniform mixing.\\
  The graph $G=(V,E)$ has the average uniform mixing property if the average probability distribution satisfies
   $P(t)=1/|V|$ for all $j\in V$}.

\subsection{The cycle graph $C_n$}

 Since the finite cycle is a regular
graph, instead of the Laplacian, we use the adjacency matrix
directly. Let $\frac{1}{2}A_n$ be the normalized adjacency matrix
of the full-cycle $C_n$ on $n$ vertices. Using the properties of
circulant matrices, the eigenvalues of $\frac{1}{2}A_n$ is given
by
\begin{equation} \label{ec}
\lambda_j=\frac{1}{2}(\omega_j-\omega_{j}^{n-1})=\cos(2\pi j/n).
\end{equation}
 The direct product for cycle graphs as
$C_{n_1}\otimes\cdots\otimes C_{n_d}$, then the corresponding
normalized adjacency matrix as follow :
 \begin{equation}
A=\frac{1}{d}\sum_{i=1}^{d} I_{n_1}\otimes\cdots\otimes
\frac{1}{2}A_{n_i}\otimes\cdots\otimes I_{n_d}
\end{equation}
where the $i$th term in the sum has $\frac{1}{2}A_{n_i}$ appearing
in the $i$th place in the tensor product. Then, we have
$$
 U_{t}=exp(-iAt)=\prod_{i=1}^{d} I_{n_1}\otimes\cdots\otimes
e^{-it\frac{1}{2d}A_{n_i}}\otimes\cdots\otimes I_{n_d}
$$
\begin{equation}
 =e^{-it\frac{1}{2d}A_{n_1}}\otimes\cdots\otimes
e^{-it\frac{1}{2d}A_{n_d}}.
 \end{equation}
 Using the eigenvalues (\ref{ec})
and orthonormal eigenvectors
$\ket{F_{j}}=\frac{1}{\sqrt{n}}\ket{\omega_{j}}$ (the columns of
the Fourier matrix F (\ref{Cir})) and the initial amplitude vector
$\ket{\psi_{0}}= \ket{0}_1 \ket{0}_2 \cdots \ket{0}_d$, then for
$\ket{\psi_{t}}$, We have
\begin{equation}
\ket{\psi_t}=U_t\ket{\psi_0}=\prod_{i=1}^{d}\otimes(\frac{1}{n_i}\sum_{j=0}^{n_i-1}e^{\frac{-it\cos(2\pi
j/n_i)}{d}}\ket{\omega_j}_i)
\end{equation}
the especially, if $n_1=n_2=\cdots=n_d$, we have
\begin{equation}
\ket{\psi_t}=U_t\ket{\psi_0}=(e^{-it\frac{1}{2d}A_{n}}\ket{0})^{\otimes
d} =(\frac{1}{n}\sum_{j=0}^{n-1}
e^{-i\lambda_{j}t/d}\ket{\omega_j})^{\otimes d}
    = (\frac{1}{n}\sum_{j=0}^{n-1}e^{\frac{-it\cos(2\pi j/n)}{d}}
    \ket{\omega_j})^{\otimes d}
\end{equation}
Thus, for calculate the amplitude of particle, we using the
amplitude the particle on the graph $G=C_n$ (i.e,
 $P_{k}(t)=\braket{k}{\psi_{t}} =
    \frac{1}{n}\sum_{j=0}^{n-1} e^{-it\cos(2\pi
    j/n)}\omega_{j}^{k}$  with  $t\longrightarrow t/d$).
    Then, the amplitude for observing particle at the position
    $\vec{K}$ is
    \begin{equation}
\braket{\vec{K}}{\psi_{t}}=\prod_{i=1}^{d}P_{k_i}(t).
    \end{equation}
 One can show only for $d=1$ with $n_1=3$ (i.e., $C_3$) and $d=2$ with  $n_1=n_2=2$  (i.e., $C_4=C_2\otimes
 C_2$) have the instantaneous exactly uniform mixing  property the
 continuous-time quantum walk model.

    \subsection{The Complete Graphs $K_{n}$}

    Let $\frac{1}{n-1}A_{n}$ be the normalized adjacency matrix of complete graph
    $K_{n}$. Thus, eigenvalues of $\frac{1}{n-1}A_{n}$ are $1$ (once) and $-\frac{1}{n-1}$ (n-1
times). The corresponding normalized adjacency matrix of direct
product complete graphs(i.e, $K_{n_1}\otimes\cdots\otimes K_{n_d}$ )
as follow :
 \begin{equation}
A=\frac{1}{d}\sum_{i=1}^{d} I_{n_1}\otimes\cdots\otimes
\frac{1}{n_i-1}A_{n_i}\otimes\cdots\otimes I_{n_d}
\end{equation}
 where the $i$th term in the sum has $\frac{1}{n_i-1}A_{n_i}$  appearing in the $i$th
place in the tensor product. Then, we have
$$
 U_{t}=exp(-iAt)=\prod_{i=1}^{d} I_{n_1}\otimes\cdots\otimes
e^{-it\frac{1}{(n_i-1)d}A_{n_i}}\otimes\cdots\otimes I_{n-d}$$
\be
=e^{-it\frac{1}{(n_1-1)d}A_{n_1}}\otimes\cdots\otimes
e^{-it\frac{1}{(n_d-1)d}A_{n_d}}.
 \ee
Using the orthonormal eigenvectors
$\ket{F_{j}}=\frac{1}{\sqrt{n}}\ket{\omega_{j}}$ (the columns of
the Fourier matrix F (\ref{Cir})) and the initial amplitude vector
$\ket{\psi_{0}}=\ket{0}_1 \cdots \ket{0}_d$, then for
$\ket{\psi_{t}}$, We have
 $$
 \ket{\psi_{t}}=U_{t}\ket{\psi_{0}}
 $$
  \be
=\prod_{i=1}^{d}\otimes \left[\frac{1}{n_i}
(e^{\frac{-it}{d}}+(n_i-1)e^{\frac{-it}{(n_i-1)d}})\ket{0}_i+\frac{1}{n_i}
(e^{\frac{-it}{d-1}}-e^{\frac{-it}{(n_i-1)d}})\sum_{l=1}^{n_i-1}\ket{l}_i\right]
 \ee
Thus the probability of particle at the position
$\ket{\vec{K}}=\ket{i_1,i_2...i_d}$, is as follow :
\begin{equation}
P_{\vec{K}}(t)=\prod_{l=1}^{d}
\frac{1}{n_l}[(e^{\frac{-it}{d}}+(n_l-1)e^{\frac{-it}{(n_l-1)d}})\delta_{i_l,0}+
(e^{\frac{-it}{d-1}}-e^{\frac{-it}{(n_l-1)d}})(1-\delta_{i_l,0})]
\end{equation}.

The especially, if $n_1=n_2=\cdots=n_d=n$ we have

  \be
 \ket{\psi_{t}}=\left[\frac{1}{n}
(e^{\frac{-it}{d}}+(n-1)e^{\frac{-it}{(n-1)d}})\ket{0}+\frac{1}{n}
(e^{\frac{-it}{d-1}}-e^{\frac{-it}{(n-1)d}})(\ket{1}+\ket{2}+\cdots+\ket{n-1})\right]^{\otimes
d }
 \ee
  and we see that the continuous-time quantum walk is
equivalent to $d$ non-interacting n-state systems. Thus the
amplitude for observing the particle at a position $\vec{k}$ is
\be \label{com} \braket{\vec{k}}{\psi_{t}}=\left[\frac{1}{n}
(e^{\frac{-it}{d}}+(n-1)e^{\frac{-it}{(n-1)d}})\right]^{k}
\left[\frac{1}{n}
(e^{\frac{-it}{d}}-e^{\frac{-it}{(n-1)d}})\right]^{d-k}
 \ee
where the $k$ is the number of zeroes. We can rewrite equation
(\ref{com}) with using the amplitude for observing the particle on
the graph $K_{n}$ \cite{abtw}  ,i.e,  if $\braket{l}{\phi_{t}}$
and $\braket{j}{\phi_{t}},$  amplitude of observing the particle
 at the site $\ket{l}=\ket{0}$ and $\ket{j}$ for
$j\neq 0$ in the time-$t$, respectively , then
\begin{equation}
\braket{\vec{k}}{\psi_{t}}=(\braket{l}{\phi_{t/d}})^{k}(\braket{j}{\phi_{t/d}})^{n-k}.
\end{equation}
One can show only for $d=1$ with $n=2,3,4$ (i.e., $K_2, K_3$ and
$K_4$) has the instantaneous exactly uniform mixing  property the
 continuous-time quantum walk model.
\subsection{Charter}
  As an example, direct product  graphs for the continuous-time quantum
   walks, we can calculate the wave function  $\ket{\psi_{t}}$ directly
   by exploiting the charter's simple structure as a product simplex
   graph $K_2$ and cycle graph $C_n$. Let
$G=Z_2\otimes Z_n$ be a complete 2-partite graph where each
partition has $n>2$ vertices(the case $n=2$ is the square graph).
The direct product of  complete graphs $K_2$ and cycle graph $C_n$
is $K_2\otimes C_n$ such that the corresponding Laplacian as follow
:
\begin{equation}
A=\frac{1}{2}(I_2\otimes\frac{1}{2}A_n+\sigma_x\otimes I_n ).
\end{equation}

If $\ket{\psi_0}=\ket{0_1}\ket{0_n}$, then, for wave amplitude
function  $\ket{\psi_t}=e^{\frac{-i\sigma_x t}{2}}\ket{0_1}\otimes
e^{\frac{-iA_{n}t}{4}}\ket{0_n}$, we have
\begin{equation}
 \braket{k}{\psi_t} = \left\{\begin{array}{ll}
            \frac{1}{n}\cos(\frac{t}{2})\sum_{j=0}^{n-1} e^{\frac{-it\cos(2\pi
    j/n)}{2}}\omega^{jk} & \mbox{ for $k=0,...,n-1$ } \\
            \frac{-i}{n}\sin(\frac{t}{2})\sum_{j=0}^{n-1} e^{\frac{-it\cos(2\pi
    j/n)}{2}}\omega^{jk}  & \mbox{for $k=n,...,2n$}
            \end{array}\right.
\end{equation}
Thus
\begin{equation}
P_k(t) = \left\{\begin{array}{ll}
            \frac{1}{n^2}\cos^{2}(\frac{t}{2})\sum_{l,j=0}^{n-1} e^{\frac{-it(\cos(2\pi
    j/n)-\cos(2\pi l/n))}{2}}\omega^{k(j-l)} & \mbox{ for $k=0,...,n-1$ } \\
            \frac{1}{n^2}\sin^{2}(\frac{t}{2})\sum_{l,j=0}^{n-1}
            e^{\frac{-it(\cos(2\pi j/n)-\cos(2\pi l/n))}{2}}\omega^{k(j-l)}  & \mbox{for $k=n,...,2n$}
            \end{array}\right.
\end{equation}
Also, for above graph's there isn't exactly uniform mixing property
under the continuous-time quantum walk model; but, for $n=3$ and
$n=4$, there is property balanced (i.e., as example for  $n=3$ at
$t=\frac{16l\pi}{3}\pm \frac{8\pi}{9}, l\in Z$  the half of
probabilities are constant and for other half  there is another
constant amount). Furthermore, there isn't average uniform mixing
property under the continuous-time quantum walk model.

  \subsection{n-Cube}
  As an example direct product simplex graph for the continuous-time quantum
   walks, we can calculate the wave function  $\ket{\psi_{t}}$ directly
   by exploiting the hypercube's simple structure as a product graph.
   The binary $n$-cube is define over the set ${(0, 1)}^n$ of
   $n$-cube binary sequence, where two sequence $x$ and $y$ are
   connected if they differ exactly in one bit position. As it
   turns out, the $n$-cube is also a complete graph.

 The corresponding normalized adjacency matrix of hypercubes as direct
 product of $K_2$  as follow:
 \be
A=\frac{1}{n}\sum_{i=1}^{n} I_{2}\otimes\cdots\otimes
\sigma_{x}\otimes\cdots\otimes I_{2} \ee where the $i$th term in
the sum has $\sigma_{x}$ (Pauli matrix) appearing in the $i$th
place in the tensor product. Then, we have
$$
 U_{t}=exp(-iAt)=\prod_{i=1}^{n} I_{2}\otimes\cdots\otimes
e^{it\sigma_{x}/n}\otimes\cdots\otimes I_{2}$$
$$
=e^{it\sigma_{x}/n}\otimes\cdots\otimes
e^{it\sigma_{x}/n}=(e^{it\sigma_{x}/n})^{\otimes n}
$$
\be
 =\left(\begin{array}{cc}cos(t/n) & i sin(t/n)  \\i sin(t/n) & cos(t/n)
 \end{array}\right)^{\otimes n}
 \ee
where $B^{\otimes n}$ is the tensor product of $n$ copies of $B$.
If $\ket{\psi_{0}}=\ket{0}^{\otimes n},$ then
 \be
\ket{\psi_{t}}=U_{t}\ket{\psi_{0}}=\left[cos(t/n)\ket{0}+i
sin(t/n)\ket{1}\right]^{\otimes n}
 \ee
 and we see that the
continuous-time quantum walk is equivalent to $n$ non-interacting
one-qubit systems. Then the amplitude for observing the particle
at a position $\vec{k}$ with Hamming weight $k$ is
 \be \label{cub}
\braket{\vec{k}}{\psi_{t}}=(cos \frac{t}{n})^{n-k}(i sin
\frac{t}{n})^k.
 \ee
We can rewrite equation (\ref{cub}) with using the amplitude for
observing the particle on the graph $G=Z_{2}$  ,i.e,  if
$\braket{0}{\phi_{t}}$ and $\braket{1}{\phi_{t}},$  amplitude of
observing for particle, respectively  at the site $\ket{0}$ and
$\ket{1}$ in the time-$t$, then
\begin{equation}
\braket{\vec{k}}{\psi_{t}}=(\braket{0}{\phi_{t/n}})^{n-k}(\braket{1}{\phi_{t/n}})^k.
\end{equation}
This implies that for $t=(2k-1)\frac{1}{\pi}n$, where $k\in
Z^{+}$, we have $P_{k}(t)=2^{-n}$, the uniform distribution.

\section{Conclusion}
We have defined direct product of graphs and have given a recipe
 for obtained probability of observing particle on vertices in the
 continuous-time classical and quantum random walk.
In the recipe  the probability of observing particle on direct
product of graphs are obtained  by multiplication of probability on
the corresponding to  sub-graphs. Also, we have shown in the
classical state the stationary uniform distribution is reached as
$t\longrightarrow \infty$ but for quantum state is not satisfy. This
recipe is useful  to determine probability of walk on complicated
graphs.  Using this method, we have calculated the probability of
continuous-time classical and quantum walk on many of finite direct
product cayley graphs ( complete cycle, complete $K_n$, charter and
$n$-cube).

\end{document}